\documentclass[acmsmall,nonacm]{acmart}
\usepackage{balance}
\usepackage[rightcaption]{sidecap}
\usepackage{xcolor}
\usepackage{latexsym}
\usepackage{tikz}
\usetikzlibrary{positioning,shapes,decorations.pathmorphing}
\usepackage{autonum} %only number referd equations.

\usepackage{microtype}
\usepackage{soul}
\usepackage{url}
\usepackage{graphicx} % DO NOT CHANGE THIS
\usepackage{multirow}
\usepackage{tabulary}
\usepackage{tabularx}
\usepackage{booktabs}
\usepackage{diagbox}
\usepackage{subcaption}
\usepackage{dsfont}
\usepackage{etoolbox}
\usepackage[linesnumbered,ruled]{algorithm2e}

\newcommand{\hide}[1]{\textcolor{gray}{#1}}
\newcommand{\up}[1]{\tiny ($\textcolor{green}{\blacktriangle}#1\%)$}
\newcommand{\down}[1]{\tiny ($\textcolor{red}{\blacktriangledown}#1\%)$}

\newcommand{\trecdl}{\textsc{TREC-DL}}
\newcommand{\trecdldf}{\textsc{Doc'19}}

\newcommand{\ms}{\textsc{MS MARCO}}
\newcommand{\beir}{\textsc{BEIR}}
\newcommand{\bert}{\textsc{BERT}}
\newcommand{\roberta}{\textsc{RoBERTa}}
\newcommand{\distilbert}{\textsc{DistilBERT}}
\newcommand{\bertl}{\textsc{BERT-Large}}

\newcommand{\bm}{\textsc{BM25}}
\newcommand{\glove}{\textsc{GloVe}}

\newcommand{\linear}{\texttt{Linear}}
\newcommand{\attention}{\texttt{Attention}}
\newcommand{\ce}{\texttt{Pointwise}}

\newcommand{\scl}{\texttt{RankingSCL}}
\newcommand{\triplet}{\texttt{RankingCTriplet}}
\newcommand{\nce}{\texttt{RankingInfoNCE}}
\newcommand{\nca}{\texttt{RankingNCA}}

\setcopyright{rightsretained}
\acmJournal{TOIS}
\begin{document}
\title{Data Augmentation for Sample Efficient and Robust Document Ranking}

\author{Abhijit Anand}
\affiliation{
  \institution{L3S Research Center}
  \city{Hannover}
  \country{Germany}
}
\email{aanand@L3S.de}

\author{Jurek Leonhardt}
\affiliation{
  \institution{Delft University of Technology}
  \city{Delft}
  \country{The Netherlands}
}
\email{L.J.Leonhardt@tudelft.nl}
\affiliation{
  \institution{L3S Research Center}
  \city{Hannover}
  \country{Germany}
}
\email{leonhardt@L3S.de}
\authornote{Research was primarily conducted while affiliated to L3S Research Center.}

\author{Jaspreet Singh}
\affiliation{
  \institution{Independent Researcher
  }
  \city{Berlin}
  \country{Germany}
}
\email{jaz7290@gmail.com}

\author{Koustav Rudra}
\affiliation{
  \institution{Indian Institute of Technology Kharagpur}
  %\city{Kharagpur}
  \country{India}
}
\email{krudra@cai.iitkgp.ac.in}
%\authornote{Research was primarily conducted while affiliated to L3S Research Center.}
\authornotemark[1]

\author{Avishek Anand}
\affiliation{
  \institution{Delft University of Technology}
  \city{Delft}
  \country{Netherlands}
}
\email{avishek.anand@tudelft.nl}
\authornotemark[1]
\begin{abstract}
Contextual ranking models have delivered impressive performance improvements over classical models in the document ranking task.
However, these highly over-parameterized models tend to be data-hungry and require large amounts of data even for fine-tuning.
In this paper, we propose data-augmentation methods for effective and robust ranking performance.
One of the key benefits of using data augmentation is in achieving \textit{sample efficiency} or learning effectively when we have only a small amount of training data.
We propose supervised and unsupervised data augmentation schemes by creating training data using parts of the relevant documents in the query-document pairs. 
We then adapt a family of contrastive losses for the document ranking task that can exploit the augmented data to learn an effective ranking model.
Our extensive experiments on subsets of the \ms{} and \trecdl{} test sets show that data augmentation, along with the ranking-adapted contrastive losses, results in performance improvements under most dataset sizes. 
Apart from sample efficiency, we conclusively show that data augmentation results in robust models when transferred to out-of-domain benchmarks.
Our performance improvements in in-domain and more prominently in out-of-domain benchmarks show that augmentation regularizes the ranking model and improves its robustness and generalization capability. 
\end{abstract}




\maketitle

\section{Introduction}
\label{sec:intro}

Recent approaches for ranking documents have focused heavily on contextual transformer-based models for both retrieval~\cite{lin2019neural,khattab2020colbert} and re-ranking~\cite{dai_sigir_2019,macavaney2019contextualized:bertir:mpi,yilmaz2019cross,hofstatter2020interpretable,li2020parade}.
To further improve the effectiveness of contextual ranking models, earlier works have explored negative sampling techniques~\cite{xiong2020approximate}, pre-training approaches~\cite{chang2020pre}, and different architectural variants \cite{khattab2020colbert, hofstatter2020interpretable}. One largely under-explored area is the use of \textit{data augmentation} in neural information retrieval (IR) to learn effective and robust ranking models.

\textit{Data augmentation} helps improve the generalization and robustness of highly-parameterized models by creating new training examples through transformations applied to the original data. 
A key benefit of data augmentation is that it can improve \textit{sample efficiency}, meaning that a model can achieve improved performance with limited amounts of training data. This is because data augmentation effectively increases the size of the training dataset, allowing a model to learn from a wider range of examples.
Additionally, data augmentation methods result in robust models allowing for better zero-shot transfer~\cite{qin2021cosda:zeroshot,riabi2021synthetic:zeroshot}.
Augmentation techniques have been successfully used to help train more robust models, particularly when using smaller datasets in computer vision~\cite{shorten_2019_image_augmentation_survey}, speech recognition~\cite{nguyen_2020_speech_aug}, spoken language understanding~\cite{peng_2020_spoken_lang}, and dialog system~\cite{zhu_2020_dialog}. 
However, the use of data augmentation for document ranking has not been investigated in detail to the best of our knowledge until recently~\cite{bonifacio2022inpars,dai2022promptagator}.

Contextual models are first \textit{pre-trained} on large amounts of language data followed by task-specific \textit{fine-tuning}. 
However, popular contextualized models are over-parameterized with more than \textit{100 million} parameters and might over-fit the training data when the task-specific fine-tuning data is small.
Many real-world ranking tasks can have smaller query workloads and therefore necessitate sample efficient training like data augmentation~\cite{muhling2022viva:small,butt2021evaluation:small,bhattacharya2019fire:small,singh2016history}.
However, simply augmenting training data with existing \textit{point-} or \textit{pairwise ranking} losses does not lead to performance improvements.
We show that our data augmentation techniques using existing pointwise ranking losses, i.e. cross-entropy losses, result in \textit{degradation of performance} (cf. Figure~\ref{fig:ce-scl}).
This can be attributed to a known lack of robustness to noisy labels~\cite{zhang2018generalized::augprob} and the possibility of poor margins~\cite{liu2016large:augprob}, leading to reduced generalization performance.

\subsection{Contrastive learning for rankings with Data Augmentation}
\label{sec:contrastive_learning}

Towards improving the ranking performance in \textit{limited data setting} we first propose both unsupervised and supervised data augmentation methods. Both cases involve creating new query-document pairs from existing instances. We do not perturb the query, only the document. 
Unsupervised data augmentation methods include adding new query document pairs where documents are relevant extractive pieces of text from an existing relevant document determined by lexical (BM25-based) or semantic (embedding-based) similarity.
For supervised augmentation, we use rationale-selection~\cite{leonhardt2021learnt} approach specifically devised for document ranking. This approach selects relevant portions of an existing relevant document in a supervised manner given a query. 

Secondly, we propose \textit{contrastive learning objectives} for document ranking that can exploit the newly augmented training instances.
A key idea in contrastive learning is to learn the input representation of an instance or \textit{anchor} such that its positive instances are embedded closer to each other, and the negative samples are farther apart.
In this work, we construct \textbf{augmented query-document} pairs from existing positive instances by multiple augmentation strategies.
We extend the idea of contrastive learning to the document ranking task by considering query-document pairs belonging to the same query as positive instances, unlike in vision and NLP tasks, where all instances with the same class label can potentially become positive pairs. 

Our key contribution to this work is the effective combination of data augmentation and contrastive learning to improve sample efficiency and robustness. 

\begin{figure}
    \centering
    \includegraphics[width=0.8\textwidth]{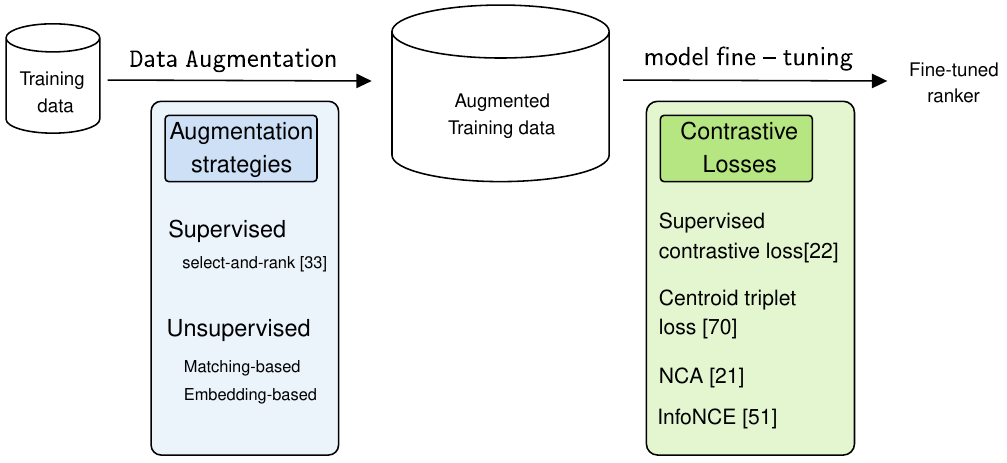}
    \caption{Training a ranking model with augmented data using different contrastive loss objectives}
    \label{fig:our_approach}
\end{figure}

\subsection{Results and Key Takeaways}

To this end, we systematically explore existing contrastive learning objectives and augmentation strategies on a host of contextual language models -- \bert{}, \roberta{} and \distilbert{} in multiple low-data ranking settings -- from $100$ to $100,000$ training instances. We do not intend to engineer a state-of-art ranking model for document ranking but instead focus on optimization strategies that work well in low-data settings. We find that using the right combination of augmentation technique and loss objective even when you have only 1k training instances leads to a $83\%$ improvement in ndcg@10 for \distilbert{}. We find that even larger models like \bert{} which tend to be more sample efficient in comparison see a 9\% improvement in low data settings. When transferring augmented models to out-of-domain datasets, we once again see drastic improvements -- \roberta{} sees sample efficiency gains ranging from $18.8\%$ to sometimes a high of $134\%$ on various \beir{} datasets with no additional fine-tuning. 

\subsection{Contributions}
\label{sec:contribution}

In sum, we make the following contributions in this work:
\begin{itemize}
    \item We propose and study different data augmentation and contrastive loss approaches for document ranking task.
    \item We also show the impact of model size on ranking performance using augmented data of different sizes.
    \item We show the performance of different data augmentation and ranking losses in in-domain and out-of-domain (\beir{}) settings.
\end{itemize}
\section{Related Work}
\label{sec:related}
The related work on the topic at hand can be broadly categorized into three main areas of research, document ranking using contextual models, different data augmentation techniques on text-based tasks, and finally, we investigate different loss functions used in text ranking and analyse their relationship with contrastive loss.

\subsection{Contextual Models for Ad-hoc Document Retrieval}
\label{sec:rel-work:contextual}

The task of text ranking often involves two steps: a fast retrieval step followed by a more involved re-ranking step. The re-ranking step is particularly important because it can significantly improve the performance of the text ranking task. In this paper, the focus is on improving the performance of the re-ranking stage which typically involves the use of contextual models.

Contextual models, such as \bert{}~\cite{devlin_bert_2018} and \roberta{}~\cite{liu2019roberta}, have shown promising improvements in the document ranking task. The input, i.e., query document pairs, can be encoded using two major paradigms for training a contextual re-ranker: joint encoding and independent encoding. In the joint encoding paradigm, the most common way to apply contextual models for document re-ranking is to jointly encode the query and document using an over-parameterized language model~\cite{nogueira_prr_2019,macavaney2019contextualized:bertir:mpi}. 
On the other hand, the second paradigm encodes the document and the query independently of each other. 
Models that implement independent query and document encoding are referred to as dual encoders, bi-encoders, or two-tower models.

While dual encoders are typically used in the retrieval phase, recent proposals have used them in the re-ranking phase as well~\cite{leonhardt2021fast}. It is important to note that a common problem in both approaches is the upper bound on the acceptable input length of contextual models, which restricts their applicability to shorter documents. When longer documents do not fit into the model, they are chunked into passages or sentences to fit within the token limit, either by using transformer-kernels~\cite{hofstatter2020local,hofstatter2020interpretable}, truncation~\cite{dai_sigir_2019}, or careful pre-selection of relevant text~\cite{rudra2020distant,leonhardt2021learnt}. As pre-training is an important part of these models, several approaches and pre-training objectives have been proposed specifically for ranking~\cite{chang2020pre,gao2021condenser,gao2022unsupervised,lassance2023experimental}.

\textbf{Ranking using LLM's}: Recently a lot of work has been done on re-ranking using LLM's in a zero-shot setup. There have been works using list-wise approach~\cite{ma2023zero,sun2023chatgpt}, Pairwise Ranking Prompting~\cite{qin2023large} for re-ranking passages. All the above approaches use either Flan-UL2 model with 20B parameters or ChatGPT (GPT-3.5-turbo,GPT-4) with more than 175B parameters. To best of our knowledge there have been no works on document ranking using LLM's in a zero-shot or instruct tuned setup.

In this work, the focus is on joint encoding models for document ranking, and simple document truncation is employed whenever longer documents exceed the overall input upper bound. By doing so, the aim is to improve the performance of the re-ranking stage and the text ranking task overall.

\subsection{Data Augmentation}
Data augmentation is a powerful technique that has a significant impact on different segments, including text, speech, image, vision, and more. Researchers have proposed new data augmentation strategies and explored their influence on deep learning models in different fields, such as speech recognition, spoken language understanding, dialog systems, image recognition, and text classification. Some of the proposed techniques include GridMask~\cite{chen_2020_gridmask}, AutoAugment~\cite{cubuk_2019_autoaugment}, and~\cite{zhong_2020_erasing,gao_2020_fuzz,raileanu_2020_generalize_reinforce,zeng_2020_adversarial,longpre_2020_effective}.

For text-related tasks, various data augmentation techniques have been proposed, such as named entity recognition, sentiment analysis, text classification, and text generation. 
Data augmentation has been shown to help boost the performance of several downstream natural language processing (NLP) and text-related tasks. For instance, data augmentation using pre-trained transformer models has been shown to improve the performance of named entity recognition, language inference, text categorization, classification, and query-based multi-document summarization tasks~\cite{kumar_2020_augment_transformer, sun_2020_mixup, shorten2021text}. 
A proposed framework called Text Attack ~\cite{morris_2020_textattack} combines data augmentation, adversarial attacks, and training in NLP. 
One of the most common data augmentation techniques in NLP is the use of word embeddings, which involves converting words into numerical vectors. Different techniques can be used to augment word embeddings, such as word substitution, word deletion, and word insertion.  Another proposed approach for text classification involves replacing words with synonyms~\cite{zhang_2021_text_augmentation} and inserting or deleting words in the input text.

In recent times, data augmentation methods have gained popularity in the context of retrieval tasks. Such techniques have shown promising results for question retrieval~\cite{nugraha_2019_typographic}, query translation~\cite{yao_2020_domain}, question-answering~\cite{yang_2019_data,yang_2020_cross_attention}, cross-language sentence selection~\cite{chen_2021_cross}, machine reading~\cite{van_2021_machine_reading}, and query expansion~\cite{lian_2020_retrieve_keyword}. For instance, Yangetal~\cite{yang2021xmoco} proposed a cross-momentum contrastive learning~\cite{he2020momentum} based scheme for open-domain question answering. Dense retriever models~\cite{xiong2020approximate,karpukhin2020dense} have recently been proposed, which sample negative documents to train the dense retrievers in a contrastive way. However, these methods do not pay attention to the uniform nature of contrastive learning~\cite{wang2020understanding}. In contrast, a contrastive dual learning based method~\cite{li_2021_dual_learning} for dense retrieval takes care of uniformity. Most of these approaches concentrate on negative samples and aim to train an effective dense retriever framework. Works like InPars~\cite{bonifacio2022inpars} and PromptAgator~\cite{dai2022promptagator} focus on using large generative models to generate queries from sampled documents to increase training data for dense retrieval tasks.

\subsection{Contrastive Learning}
\label{sec:rel-work:scl}
Contrastive losses with data augmentation have been widely studied in unsupervised learning settings, with augmentation of instances treated as positive samples and other random instances serving as negative samples. However, recent research has explored ways to incorporate label information for more precise supervision signals during data augmentation~\cite{khosla2020supervised}. Various methods have used this approach to learn representations from unsupervised data~\cite{oord2018representation,hjelm2018learning,wu2018unsupervised,sohn2016improved}, achieving superior performance compared to other approaches~\cite{donahue2019large,gidaris2018unsupervised}. By generating training instances from original ones using different data augmentation strategies, a contrastive loss can help bring the representation of related entities closer together in the embedding space. For a comprehensive overview, a recent survey on supervised and self-supervised contrastive learning is recommended~\cite{jaiswal2021survey}.
Recent research has applied supervised contrastive learning (SCL) to fine-tuning pre-trained language models, but with limited success~\cite{gunel2020supervised}. 
There are various other contrastive losses used in text retrieval, which include InfoNCE loss\cite{oord2018representation}, N-Pair loss~\cite{sohn2016improved}, centroid triplet loss~\cite{wieczorek2021unreasonable}, and lifted structured loss~\cite{sohn2016improved}. 
We discuss in detail the contrastive losses in Section~\ref{sec:contrastive-learning}.

Another line of work that frequently utilizes contrastive loss functions is \emph{dense passage retrieval}. \citet{karpukhin2020dense} introduced the DPR model, which uses dual-encoders to independently compute representations of queries and documents. The loss used to train DPR models is similar to the ones mentioned above; it makes use of \emph{in-batch negatives}, i.e., it takes the documents from \emph{all} instances in a given training batch into account and \emph{contrasts} them with the relevant document. Several other approaches~\cite{zhu2021contrastive,qu2021rocketqa} have adopted this training objective. Subsequent works have focused on providing better \emph{negative sampling} techniques to replace the in-batch negatives~\cite{xiong2020approximate,lindgren2021efficient,zhan2021optimizing}. In principle, our data augmentation techniques can be applied in the context of dual-encoder models for retrieval, although we focus on cross-encoders for re-ranking in this work.

\subsection{Extension from previous work}
This work is an extension of our previous work~\cite{anand2022supervised} titled ``Supervised Contrastive Learning Approach for Contextual Ranking''.
In~\cite{anand2022supervised} we predominantly explored simple data augmentation techniques with supervised contrastive loss (SCL) for document ranking.
In this work, we considerably increase the scope of our investigations to include more involved supervised data augmentation schemes like~\cite{leonhardt2021learnt}.
Also included in this paper is a detailed investigation of other metric losses like centroid triplet loss, noise contrastive losses, and neighborhood-based unsupervised losses.
Different from~\cite{anand2022supervised}, we also study how model size impacts model performance when trained on different sizes of augmented and non-augmented data.
Finally, we showcase the benefits of data augmentation in zero-shot transfer settings to test the robustness and generalization of the rankers learned using augmented training data.
\section{Document (Re-)Ranking using Contextual Language Models}
\label{sec:problem}
Our task is to train a model for \textit{document re-ranking}. Ranking models usually provide a relevance score when given a query-document pair $(q, d)$ as input. This score can then be used to rank documents based on their relevance to the given query.

Formally, the training set comprises pairs ${q_i, d_i }_{i=1}^{N}$, where $q_i$ is a query and $d_i$ is a document that is either relevant or irrelevant based on its label $y_i$. The aim is to train a ranker $R$ that predicts a relevance score $\hat{y} \in [0; 1]$ given a query $q$ and a document $d$:
\begin{equation}
R: (q, d) \mapsto \hat{y}
\end{equation}
After training, the ranking model $R$ can be used to re-rank a set of documents obtained in the initial retrieval process by a lightweight, typically term-frequency-based, retriever with respect to a query. This is a common practice for ranking tasks, where the documents are initially retrieved and then re-ranked by a more sophisticated and computationally expensive model. Recent studies have shown that pre-trained contextual language models have exhibited promising performance in document ranking tasks~\cite{nogueira_prr_2019,dai_sigir_2019,rudra2020distant,yilmaz2019cross}. These cross-attention models jointly model queries and documents. In this study, three different joint modeling approaches based on \bert{}\cite{devlin_bert_2018}, \roberta{}\cite{liu2019roberta} and \distilbert{}~\cite{sanh2019distilbert} are considered, and their performance is evaluated under different contrastive loss setup with different amounts of data augmentation. All three models share the same input format: a pair of query $q$ and document $d$ is fed into the model as
\begin{equation}
\texttt{[CLS]}\ q\ \texttt{[SEP]}\ d\ \texttt{[SEP]}
\end{equation}
To account for the input length limitations of the models, long documents may need to be truncated to fit the sequence length the model is pre-trained with.

Traditionally, there are two main methods to train ranking models, which are pointwise and pairwise. Let us assume of a mini batch of $N$ training examples $\left\{x_i, y_i\right\}_{i=1, ..., N}$. The pointwise training method considers the document ranking task as a binary classification problem, where each training instance $x_i = (q_i, d_i)$ is a query-document pair and $y_i \in {0, 1}$ is a relevance label. The predicted score of $x_i$ is denoted as $\hat{y}i$. The cross-entropy loss function is defined as follows:
\begin{equation}
    \mathcal{L}_{\mathtt{Point}} = -\frac{1}{N} \sum_{i=1}^{N} \left( y_{i} \cdot \log \hat{y}_{i} + (1 - y_{i}) \cdot \log (1-\hat{y}_{i}) \right)
\end{equation}

In the pairwise training method, each training example contains a query and two documents, $x_i = (q_i, d^+i, d^-i)$, where the former is more relevant to the query than the latter. The pairwise loss function is defined as follows:
\begin{equation}
    \mathcal{L}_{\mathtt{Pair}} = \frac{1}{N} \sum_{i=1}^{N} \max \left\{0, m - \hat{y}^+_i + \hat{y}^-_i \right\}
\end{equation}
where $\hat{y}^+_i$ and $\hat{y}^-_i$ are the predicted scores of $d^+_i$ and $d^-_i$, respectively, and $m$ is the loss margin. The pairwise method is commonly used for ranking tasks as it takes into account the relative ordering between documents.

\section{Data Augmentation for Document Ranking}
\label{sec:augmentation}

We intend to use data augmentation in the context of a document ranking task to improve the quality of the training data and increase the diversity of the examples presented to the model. 
In this section, we propose extractive methods to create new query-document pairs from the instances already in the training data set.

To enhance the training data, we utilize augmentation techniques to form $d_a^+$ from $d^+$ for each triple $(q, d^+, d^-)$ in the training set. Creating an augmented instance is extractive since it involves \textit{selecting} relevant sentences to the corresponding query, followed by random sampling of an irrelevant document $d_a^-$. The resulting augmented training instances are then appended to their respective batch, effectively doubling the size of each batch. 

The document is treated as a sequence of sentences $s_i$, denoted as $d = (s_1, s_2, ..., s_{|d|})$. A query-specific selector is employed to choose a fixed number of sentences from the document, based on the distribution $p(s\mid q, d)$, encoding the relevance of the sentence given the input query $q$. This distribution is used to select an extractive, query-dependent summary, denoted as $d' \subseteq d$. 
The augmentation process is detailed in Algorithm~\ref{alg:augmentation}. The function that creates the augmented document (line~\ref{alg:augmentation:augment}) is defined as
\begin{equation}
    \label{eqn:augmentation}
    \texttt{augment}(d, q, k) = k \text{-argmax}_{1 \leq i \leq |d|} \operatorname{score}(q, s_i)
\end{equation}
The sentences in the augmented documents are ordered by score, i.e., the original order is not preserved. Note that the scoring function $\operatorname{score}(q, s)$ in Eq.~\eqref{eqn:augmentation} represents an augmentation strategy. We present both unsupervised (heuristic) and supervised (predictive) augmentation strategies in the following sections.

\begin{algorithm}[t]
    \DontPrintSemicolon
    \SetKwFunction{Augment}{augment}
    \SetKw{In}{in}
    \KwIn{training batch $B$, number of sentences per augmented document $k_a$}
    \KwOut{augmented training batch $B'$}
    $B' \leftarrow$ empty list\;
    \ForEach{$(q, d^+, d^-)$ \In $B$}{
        \tcp{keep the original example}
        append $(q, d^+, d^-)$ to $B'$\;
        \tcp{create augmented example}
        $d^+_a \leftarrow \Augment(d^+, q, k_a)$\; \label{alg:augmentation:augment}
        $d^-_a \leftarrow$ random irrelevant document\;
        append $(q, d^+_a, d^-_a)$ to $B'$\;
    }
    \Return{$B'$}\;
    \caption{Training data augmentation}
    \label{alg:augmentation}
\end{algorithm}

\subsection{Unsupervised augmentation} 
\label{sec:unsup-augmentation}

\subsubsection{Term-matching-based (\bm{})}
BM25 (Best Matching 25) is a variant of the popular tf-idf weighting scheme that is commonly used to rank documents by relevance to a query. \bm{} works by computing a score for each document in a corpus based on its relevance to a query. The score is calculated by combining several factors, including the frequency of the query terms in the document, the inverse document frequency of the query terms, and the length of the document, and is then normalized by the average document length in the corpus. We use tf-idf scores between the query $q$ and sentences $s_i$ to determine the best sentences:
\begin{equation}
    \operatorname{score}_\bm(q, s_i) = \bm(q, s_i).
\end{equation}
Inverse document frequencies are computed over the complete corpus.

\subsubsection{Semantic-similarity-based (\glove{})}
\glove{} works by constructing a co-occurrence matrix of word pairs based on their co-occurrence frequency in a corpus. The co-occurrence matrix is then factorized using matrix factorization techniques, such as singular value decomposition (SVD), to generate low-dimensional embeddings for each word. Unlike other word embedding methods, such as Word2Vec, \glove{} is designed to capture both the global co-occurrence statistics and the local context of the words. It does this by weighting the importance of word co-occurrences based on their frequency and using a logarithmic function to down weight the importance of highly frequent co-occurrences. We use \glove{} to find the representation for query $q$ and sentence $s_i$. Both the query and sentence are represented as average over their constituent word embeddings. We use semantic (cosine) similarity scores between the query $q$ and sentences $s_i$ to determine the best sentences for a given query, i.e.,
\begin{equation}
    \operatorname{score}_\glove(q, s_i) = \langle \operatorname{E}_\glove(q), \operatorname{E}_\glove(s_i) \rangle,
\end{equation}
where $\langle \cdot, \cdot \rangle$ is the dot product and $\operatorname{E}_\glove(\cdot)$ computes the average of the embeddings of all tokens in a sequence.

\subsection{Supervised Augmentation} 
\label{sec:sup-augmentation}

Unlike unsupervised augmentation where sentence selections were based on lexical- and semantic- similarities, we also consider sentence selections based on supervised training data.
Recently, in the area of explainable information retrieval, select-and-rank approaches have been proposed that given a query-document pair, select an extractive piece of text from the document as a potentially relevant signal~\cite{zhang2021explain,leonhardt2021learnt,li2023power}. 
Specifically, in the selection phase, relevant sentences given a query is extracted. 
This is followed by the ranking phase, where the relevance estimation is performed just on the extracted sentences. 
The key idea here is that typically a small part of the document is relevant and the selection phase filters out non-relevant text. 
The supervision signal is obtained by the training data and a combination of the gumbel-max trick and reservoir sampling is used to train the selector network~\cite{leonhardt2021learnt}.
The output of the selection phase can be considered as a query-based extractive summary and can be utilized for our data augmentation needs.
We considered the two supervised sentence selection approaches -- \linear{} sentence selection, and \attention{} based sentence selection.

Let a query $q = \left( t^q_1, ..., t^q_{|q|} \right)$ and document $d = \left( t^d_1, ..., t^d_{|d|} \right)$ be sequences of (embedded) tokens, respectively. Furthermore, let $s_{ij}$ be a sentence within the document, such that $s_{ij} = \left( t^d_i, ..., t^d_j \right)$. In the following, we describe how the two selection approaches score a sentence $s_{ij}$ w.r.t.\ a query $q$.

\subsubsection{Linear sentence selection}
The linear sentence selector is a simple non-contextual model, i.e., each sentence within the document is scored independently. This approach is similar to the \glove{}-based augmentation, however, this model has been trained specifically on a ranking task. The query and each sentence are first represented as the average of their token embeddings, respectively. After averaging, each representation is passed through a single-layer feed-forward network. The final score of a sentence is then computed as the dot product of its representation and the query representation. Formally, a sequence of tokens, $t = \left( t_1, ..., t_{|t|} \right)$, is represented as
\begin{equation}
    \operatorname{Enc}(t) = \frac{\sum_{s_i \in t} (W t_i + b)}{|t|},
\end{equation}
where $W$ and $b$ are trainable parameters of the feed-forward layer. The score of a sentence $s_{ij}$ w.r.t.\ the query $q$ is then computed as
\begin{equation}
    \operatorname{score_{Lin}}(q, s_{ij}) = \langle \operatorname{Enc}(q), \operatorname{Enc}(s_{ij}) \rangle,
\end{equation}
where $\langle \cdot, \cdot \rangle$ is the dot product.

\subsubsection{Attention-based sentence selection}
The Attention-based selector computes sentence-level representations based on the \textsc{QA-LSTM} model~\cite{tan2016improved}. Query and document are first contextualized by passing their token embeddings through a shared bi-directional LSTM:
\begin{align}
    q^\text{LSTM} &= \operatorname{Bi-LSTM}(q), \\
    d^\text{LSTM} &= \operatorname{Bi-LSTM}(d).
\end{align}
The query representation $\hat{q}$ is obtained by applying element-wise max-pooling over $q^\text{LSTM}$:
\begin{equation}
    \hat{q} = \operatorname{Max-Pool}(q^\text{LSTM})
\end{equation}
For each hidden representation $d^\text{LSTM}_i$, attention to the query is computed as
\begin{align}
    m_i &= W_1 h_1 + W_2 \hat{q}, \\
    h_i &= d^\text{LSTM}_i \exp \left( W_3 \tanh \left( m_i \right) \right),
\end{align}
where $W_1$, $W_2$ and $W_3$ are trainable parameters. For $s_{ij}$, let $h_{ij}$ denote the corresponding attention outputs. The sentence representation is computed similarly to the query representation, i.e.,
\begin{equation}
    \hat{s}_{ij} = \operatorname{Max-Pool}(h_{ij}).
\end{equation}
The final score of a sentence is computed as the cosine similarity of its representation and the query representation:
\begin{equation}
    \operatorname{score_{Att}}(q, s_{ij}) = \operatorname{cos}(\hat{q}, \hat{s}_{ij}).
\end{equation}

\subsection{Creating Augmented Training Batches}
\label{sec:batch_creation}
To preserve the randomness in the data while augmenting the training set, the creation of mini-batches is a crucial step in our approach. It has been demonstrated in prior studies that the quality of the augmented data plays an important role in the performance of self-supervised contrastive learning~\cite{gunel2020supervised}.

Our approach begins with retrieving the top-$k$ documents per query using a first stage retrieval method. From this top-$k$ set, we create the training dataset by collecting all positive query-document training instances. For each positive pair, we randomly sample one irrelevant document to serve as the negative instance. To ensure randomness, we shuffle the resulting set of $(q, d^+, d^-)$ triples. It should be noted that, for pointwise training, we create two query-document pairs from each triple.
\section{Contrastive Learning with Augmented Data for Document Ranking}
\label{sec:contrastive-learning}

Augmented training data can be used to train a model with no alteration to the pointwise and pairwise cross entropy loss objective. We instead propose the usage of contrastive learning to better leverage augmented data. Augmented query document pairs are not clean training samples by definition and should be treated accordingly during training.

In this section, we detail different contrastive learning objectives which we used to train our ranking models. We do not propose a new contrastive loss objective but instead focus on combining existing losses correctly for ranking tasks. 

\begin{figure}
    \centering
    \includegraphics[width=1.0\textwidth]{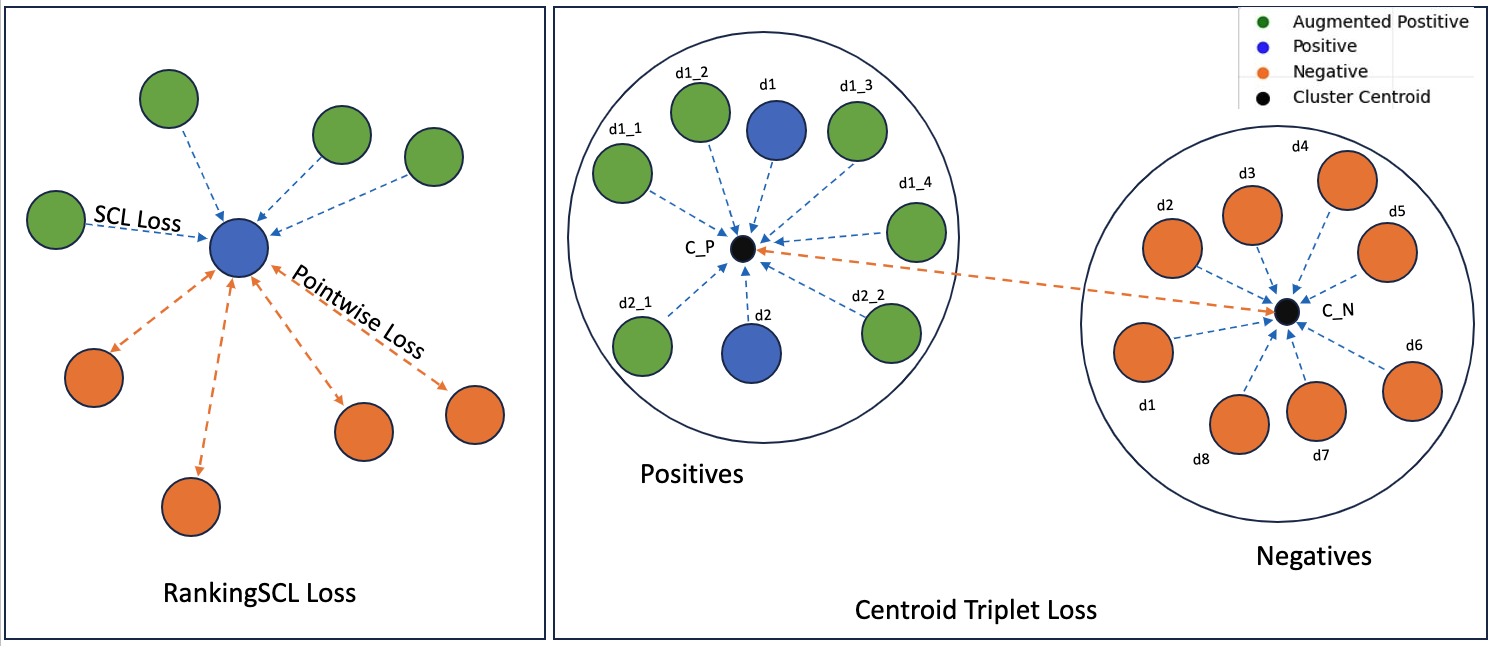}
    \caption{On the left we describe RankingSCL loss (Eq. \eqref{eqn:scl_equation}) and on the right Centroid triplet Loss (Eq. \eqref{eqn:ctl_loss})}
    \label{fig:scl_ctl_loss}
\end{figure}
\subsection{Ranking-based Supervised Contrastive loss}
\label{sbsec:scl_loss}

The aim of the Supervised Contrastive loss (SCL loss) is to capture the similarities between relevant document parts for a given query and contrast them against examples from non-relevant queries. The ranking model outputs the query-document representation $\Phi(\cdot) \in \mathbb{R}^{t}$ (e.g., the \texttt{[CLS]} output for \bert{}-based models) which can be used to compute similarities. The SCL loss function includes the adjustable scalar temperature parameter $\tau>0$ that controls the distance between relevant and non-relevant examples, and the scalar weighting hyper-parameter $\lambda$, which is tuned for each downstream task and setting. 

The SCL loss is formulated as
\begin{equation}
\label{eqn:scl_loss}
    \mathcal{L}_{\mathtt{SCL}} = 
        \sum_{i=1}^{N}-\frac{1}{N_{+}}
        \sum_{j=1}^{N_{+}}
        \mathbf{1}_{\substack{q_i = q_j, \\
                    i \neq j, \\
                    y_i = y_j = 1}}
        \log \frac  {\exp \left(\Phi \left(x_{i} \right) \cdot \Phi \left(x_{j} \right) / \tau \right)}
                    {\sum_{k=1}^{N}
                        \mathbf{1}_{i \neq k} \exp \left(\Phi \left(x_{i} \right) \cdot \Phi \left(x_{k} \right) / \tau \right)}
\end{equation}

where $N_{+}$ is the total number of positive examples (i.e., relevant query-document pairs) in the batch. 

The loss function enforces that the positive pair with the same query should be embedded close to each other, rather than a pair of documents that are relevant for different queries. This is important since it ensures that the representations for the "relevant parts" of the same query are close to each other. The final ranking SCL loss is

\begin{equation}
\label{eqn:scl_equation}
\mathcal{L}_{\mathtt{RankingSCL}} = (1-\lambda) \mathcal{L}_{\mathtt{Ranking}} + \lambda \mathcal{L}_{\mathtt{SCL}}
\end{equation}

We illustrate the \scl{} loss in Figure \ref{fig:scl_ctl_loss} (left figure) using a pointwise ranking loss. It shows the two components working together; the ranking loss separates the pairs of positive and negative documents, while the contrastive loss moves all positive documents in the batch closer to each other.

\subsection{Ranking-based Centroid Triplet Loss}
\label{sec:centroid-triplet-loss}

The triplet loss function~\cite{wieczorek2021unreasonable} enforces the learning of a distance metric that satisfies the following property: for a given triplet of data points ($A$, $P$, $N$), where $A$ is an anchor point (same class), $P$ is a positive example for A, and $N$ is a negative example for $A$, the distance between $A$ and $P$ should be smaller than the distance between $A$ and $N$ by a certain margin. This property is known as the triplet constraint. The objective is to minimize the distance between $A-P$, while maximizing the distance to the $N$ sample.

The loss function is formulated as follows:
\begin{equation}
\mathcal{L}_{\text {triplet }}=\left[\|f(A)-f(P)\|_2^2-\|f(A)-f(N)\|_2^2+\alpha\right]_{+}
\end{equation}
where $[z]_{+}=\max (z, 0), f$ denotes embedding function learned during training stage and $\alpha$ is a margin parameter.

In the case of Centroid Triplet Loss (CTL), instead of comparing the distance of an anchor $A$ to the positive and negative instances, CTL measures the distance between $A$ and class centroids $c_P$ and $c_N$ representing either the same class as the anchor or a different class respectively.
In case of a document ranking task, for a given query $q$, let $\left\{{d^+}_i, {d^-}_i\right\}_{i=1, ..., N}$ be a set of relevant and non-relevant documents respectively to that given query. Where ${d^+}_i$ can be positives or augmented positives. Let $c_P$ be the centroid of the relevant class (relevant and augmented relevant documents) and $c_N$ be centroid of the non-relevant class. Then CTL is therefore formulated as:

\begin{equation}
\label{eqn:ctl_loss}
\mathcal{L}_{\text {CTriplet }}=\left[\left\|d^+_i -c_P\right\|_2^2-\| d^-_i -c_N\|_2^2+\alpha_c\right]_{+}
\end{equation}

\begin{equation}
\label{eqn:ctl_equation}
\mathcal{L}_{\mathtt{RankingCTriplet}} = (1-\lambda) \mathcal{L}_{\mathtt{Ranking}} + \lambda \mathcal{L}_{\mathtt{triplet}}
\end{equation}

\subsection{Ranking-based InfoNCE}
\label{sec:info-nce}

The InfoNCE loss function\cite{oord2018representation} encourages similar samples to have similar representations and dissimilar samples to have dissimilar representations. This is achieved by comparing the similarity between positive pairs (similar samples) and negative pairs (dissimilar samples) using the cross-entropy loss. The InfoNCE loss function is a variant of the standard cross-entropy loss that has been modified to account for the varying number of negative samples used in the contrastive learning framework.

InfoNCE is particularly effective in situations where the number of negative samples is large. Additionally, the use of the cross-entropy loss makes the InfoNCE loss easy to optimize using standard optimization algorithms.

In document ranking given a query $\mathbf{q_i}$ and corresponding relevant document $\mathbf{d^+_i}$, the positive sample should be drawn from the conditional distribution $p(\mathbf{x} \mid \mathbf{d^+_i})$, while $N-1$ negative samples are drawn from the proposal distribution $p(\mathbf{x})$, independent from the context $d^+_i$ . For simplicity, let us label all the documents for the query $q_i$ as $D=\left\{\mathbf{d}_i\right\}_{i=1}^N$ among which only one of them $\mathbf{d}_{\text {pos}}$ is a positive sample. The probability of we detecting the positive sample correctly is:
\begin{equation}
p(C=\operatorname{pos} \mid D, \mathbf{d^+_i})=\frac{p\left(d_{\mathrm{pos}} \mid \mathbf{d^+_i}\right) \prod_{i=1, \ldots, N ; i \neq \mathrm{pos}} p\left(\mathbf{x}_i\right)}{\sum_{j=1}^N\left[p\left(\mathbf{x}_j \mid \mathbf{d^+_i}\right) \prod_{i=1, \ldots, N ; i \neq j} p\left(\mathbf{x}_i\right)\right]}=\frac{\frac{p\left(\mathbf{x}_{\mathrm{pos}} \mid d^+_i\right)}{p\left(\mathbf{x}_{\mathrm{pos}}\right)}}{\sum_{j=1}^N \frac{p\left(\mathbf{x}_j \mid \mathbf{d^+_i}\right)}{p\left(\mathbf{x}_j\right)}}=\frac{f\left(\mathbf{x}_{\mathrm{pos}}, \mathbf{d^+_i}\right)}{\sum_{j=1}^N f\left(\mathbf{x}_j, \mathbf{d^+_i}\right)}
 \end{equation}
 
where the scoring function is $f(\mathbf{x}, \mathbf{d^+_i}) \propto \frac{p(\mathbf{x} \mid \mathbf{d^+_i})}{p(\mathbf{x})}$.
The InfoNCE loss optimizes the negative log probability of classifying the positive sample correctly:
\begin{equation}
\label{eqn:infonce_loss}
\mathcal{L}_{\text {InfoNCE }}=-\mathbb{E}\left[\log \frac{f(\mathbf{x}, \mathbf{d^+_i})}{\sum_{\mathbf{x}^{\prime} \in X} f\left(\mathbf{x}^{\prime}, \mathbf{d^+_i}\right)}\right]
\end{equation}

The fact that $f(x, d^+_i)$ estimates the density ratio $\frac{p(x \mid d^+_i)}{p(x)}$ has a connection with mutual information optimization. To maximize the the mutual information between input $x$ and context vector $d^+_i$, we have:
\begin{equation}
I(\mathbf{x} ; \mathbf{d^+_i})=\sum_{\mathbf{x}, \mathbf{d^+_i}} p(\mathbf{x}, \mathbf{d^+_i}) \log \frac{p(\mathbf{x}, \mathbf{d^+_i})}{p(\mathbf{x}) p(\mathbf{d^+_i})}=\sum_{\mathbf{x}, \mathbf{d^+_i}} p(\mathbf{x}, \mathbf{d^+_i}) \log {\frac{p(\mathbf{x} \mid \mathbf{d^+_i})}{p(\mathbf{x})}}
\end{equation}
where the logarithmic term in below is estimated by $f$.
\begin{equation}
    {\frac{p(\mathbf{x} \mid \mathbf{d^+_i})}{p(\mathbf{x})}}
\end{equation}

\begin{equation}
\label{eqn:infonce_equation}
\mathcal{L}_{\mathtt{RankingInfoNCE}} = (1-\lambda) \mathcal{L}_{\mathtt{Ranking}} + \lambda \mathcal{L}_{\mathtt{InfoNCE}}
\end{equation}

\subsection{Neighbourhood Component Analysis (NCA)} 
\label{subsec:NCA_loss}
NCA~\cite{goldberger2004neighbourhood} is a distance-based classification algorithm that learns a metric for measuring the similarity between data points. The metric is learned based on a training set of labeled examples and is used to classify new, unseen examples.
The basic idea behind NCA is to learn a linear transformation of the input data that maximizes the accuracy of a k-Nearest Neighbor (k-NN) classifier. The transformation is learned by minimizing a loss function that measures the classification error of the k-NN classifier.
In context of ranking, given a query $q_i$ and corresponding relevant documents $d_i$ and $d_j$, the NCA loss function is defined as follows:

\begin{equation}
\label{eqn:nca_loss}
\mathcal{L}_{\mathtt{NCA}}=\sum_i \log \left(\sum_{j \in C_i} p_{i j}\right)=\sum_i \log \left(p_i\right)
\end{equation}

where $p_{i}$ is the probability of calculating the document $d_i$ to the relevant class as neighbouring point $d_j$ is defined as:

\begin{equation}
p_i=\sum_{j \in C_i} p_{i j}
\end{equation}

We define the $p_{i j}$ using a softmax over Euclidean distances in the transformed space:

\begin{equation}
p_{i j}=\frac{\exp \left(-\left\|A d_i-A d_j\right\|^2\right)}{\sum_{k \neq i} \exp \left(-\left\|A d_i-A d_k\right\|^2\right)}
\end{equation}

where A is the transformation matrix, $d_i$ is relevant to the query $q_i$ and $d_k$ is the k-Nearest relevant Neighbor of $d_i$ in the input space.

\begin{equation}
\label{eqn:nca_equation}
\mathcal{L}_{\mathtt{RankingNCA}} = (1-\lambda) \mathcal{L}_{\mathtt{Ranking}} + \lambda \mathcal{L}_{\mathtt{NCA}}
\end{equation}

\subsection{Combining Ranking Loss with Contrastive Loss}
We propose a simple linear combination of standard ranking losses with contrastive losses described above for the augmented training samples. The overall ranking losses are then given by

\begin{equation}
\mathcal{L}_{\mathtt{ContrastiveRanking}} = (1-\lambda) \mathcal{L}_{\mathtt{Ranking}} + \lambda \mathcal{L}_{\mathtt{Contrastive}},
\end{equation}
where
\begin{align}
    \mathcal{L}_{\mathtt{Ranking}} &\in \{\mathcal{L}_{\mathtt{Point}}, \mathcal{L}_{\mathtt{Pair}}\} \text{ and}\\
    \mathcal{L}_{\mathtt{Contrastive}}
    &\in 
    \{\mathcal{L}_{\mathtt{SCL}},
    \mathcal{L}_{\mathtt{triplet}},
    \mathcal{L}_{\mathtt{InfoNCE}},
    \mathcal{L}_{\mathtt{NCA}}\}.
\end{align}
We use the following terminology in the paper: linear interpolation of \ce{} and SCL is referred to as \textbf{\scl{}}.
Although all of the aforementioned losses are contrastive, each of them has subtle differences in the way they optimize learning of the representation space.

\begin{itemize}
    \item  \textit{Supervised Contrastive Loss} (Eq.\ref{eqn:scl_loss}) encourages the features of positive examples from the same class to be similar while making sure that the features of negative examples from different classes are dissimilar. 

    \item \textit{Centroid Triplet Loss} (Eq.\ref{eqn:ctl_loss}) is a contrastive loss function that is designed to encourage a larger \textit{margin} between the distances of the positive and negative examples compared to the distance between the anchor and the centroid of the positive examples. This means that the loss function places a greater emphasis on making sure that the positive examples are tightly clustered around their centroid, while the negative examples are kept farther away. 

    \item \textit{Neighborhood Component Analysis (NCA)} (Eq.\ref{eqn:nca_loss})  is a metric learning algorithm that is designed to learn a linear transformation that maximizes the accuracy of the k-nearest neighbors (KNN) classifier. The goal of NCA is to find a transformation that reduces the distance between examples that belong to the same class and increases the distance between examples from different classes. 

    \item \textit{InfoNCE (InfoMax Contrastive Estimation)} (Eq.\ref{eqn:infonce_loss}) is based on the concept of maximizing mutual information between positive examples and minimizing it between negative examples. It is often used in self-supervised learning tasks where there is no labeled data available.
\end{itemize}

In summary, the centroid triplet loss and 
supervised contrastive Loss is used for supervised learning tasks where labeled data is available, while NCA and InfoNCE can be used for unsupervised or self-supervised learning tasks where labeled data is not available.
\section{Experimental Setup}
In this section, we describe the setup we used to answer the research questions. Note that we focus on the re-ranking task and not the retrieval task. 

\subsection{Datasets}
We conduct experiments on in-domain and out-of-domain benchmarks to showcase the utility and performance benefits of data augmentation.

\subsubsection{In-Domain Benchmark - \trecdl{}} In this study, we utilize the dataset provided by the TREC Deep Learning track in 2019. We evaluate our proposed model on \trecdldf{}, comprising of 200 distinct queries. The training and development set is obtained from \ms{}, which contains a total of 367K queries. For the retrieval of the top 100 documents for each query, we use Indri~\cite{msmarco_trec_2019}.

\subsubsection{Out of Domain Benchmark - \beir{}}
The \beir{} dataset is a collection of datasets used for benchmarking and evaluating the performance of information retrieval (IR) models~\cite{thakur2021beir}. 
The datasets are selected from various domains, such as news articles, scientific papers, and product reviews. It consists of $17$ different datasets.

\subsection{Ranking Models}
We use different cross-attention models for our experiments:

\begin{enumerate}
    \item \textbf{\bert{}} \cite{devlin_bert_2018}, a transformer-based large, pre-trained contextual model. Our implementation employs the base version of BERT, which is composed of 12 encoder layers, 12 attention heads, and 768-dimensional output representations. Additionally, the maximum input length is limited to 512 tokens.
    
    \item \textbf{\roberta{}} \cite{liu2019roberta} is another cross-attention model which is architecturally identical to \bert{}; the only difference between the two models is the pre-training procedure. 
    
    \item \textbf{\distilbert{}} \cite{sanh2019distilbert} is a smaller, distilled version of \bert{} that is designed for faster and more efficient training and inference. It has only 6 transformer layers, 768-dimensional output representations, and a maximum input length of 512 tokens. \distilbert{} was trained using a novel technique called knowledge distillation, where a larger pre-trained model is used to teach a smaller model, resulting in significant reductions in model size and computational cost while maintaining a high level of performance.
    
    \item \textbf{\bertl{}} \cite{devlin_bert_2018} is a large-scale pre-trained language model that uses a bidirectional transformer architecture to learn representations of text. It has 336 million parameters, which is 4 times larger than the original BERT-base model. It has 24 encoder layers, 16 attention heads, and 1024-dimensional output representations.
\end{enumerate}

We experimented with (a) different types of contextual models – \bert{}, \roberta{}, \distilbert{}, (b) varying dataset sizes – {100, 1K, 2K, 10K, 100K} instances for \trecdldf{} (c) ranking losses – \scl{}, \nce{}, \triplet{}, \nca{} d) different datasets \trecdldf{}, \beir{}. This leads to a large experimental space for exploration which we have synthesized into key insights in our results. For instance, the number of models trained on \trecdldf{} is around 1050 with 208 best combinations chosen for reporting. 
 Given a large number of models, it is difficult to report all combination of results and their respective hyperparameters. So we report the best models in the paper and a subset of the results, hyperparameters and  can be found in Appendix ~\ref{appendix:appendix}.

\subsection{Batch Creation and Hyperparameters}
\label{sec:batch-creation}

In Section~\ref{sec:batch_creation}, we described our approach to creating batches for supervised contrastive learning, where we start with positive query-document pairs from the top-$k$ retrieved set and randomly sample negative pairs for the original dataset. We also use a selector to generate augmented versions of documents. For evaluating our approach, we experimented with different sizes of query-document pairs for \ms{}, including 1k, 2k, 10k, and 100k. For instance, the 1k dataset contains 500 positive and 500 negative pairs in the original dataset, to which we add 1k more pairs through the augmentation process, resulting in a total of 2k query-document pairs. This pattern holds for the other three sizes as well. It is worth noting that we only augment the training data, and the validation and test sets remain unaltered.

\paragraph{\textbf{Hyperparameters}}
We have two hyperparameters in our models with \scl{}: the temperature ($\tau$), and the degree of interpolation ($\lambda$) as in \scl{} [Eq. \eqref{eqn:scl_equation}]. For all other losses ([Eq. \eqref{eqn:ctl_equation}][Eq. \eqref{eqn:infonce_equation}][Eq. \eqref{eqn:nca_equation}]) we only have one hyperparameter ($\lambda$) for interpolation.
We use the \ms{} development set to determine the best combination of hyperparameters. These parameters are different for different ranking models and augmentation strategies. For example, in \trecdl{}, \bert{} ranking model using \attention{} data augmentation and \triplet{} loss objective returns the best score on the validation set at $\lambda=0.3$. In all our experiments we use a batch size of 16. A brief of the hyperparameters used is given in Appendix ~\ref{appendix:hyperparameters}.

\section{Experimental Results}
\label{sec:results}
The results section is divided into 3 major subsections. Section~\ref{sec:in-domain-exp} presents the results of our in-domain experiments where we strictly observe how augmentation and contrastive learning can benefit sample efficiency and performance for the same dataset. Then we discuss the out-of-domain performance of the models born out of our training regimen. More specifically, we study the zero-shot transfer of our models to other ranking datasets that vary in topicality and type of query in Section~\ref{sec:out-domain-exp}. Finally, we discuss the failure cases, limitations, and drawbacks of our approach in Section~\ref{sec:fail}
The key research questions we seek answers for are as follows:

The key research questions we seek answers for are as follows:
\begin{itemize}
    \item[\textbf{RQ I.}] Does data augmentation require a different training paradigm?
    \item[\textbf{RQ II.}]  Which data augmentation technique provides the highest sample efficiency gains?
    \item[\textbf{RQ III.}]  Which contrastive loss objective leads to the highest gains when data augmentation type and budget are fixed?
    \item[\textbf{RQ IV.}] Does our training regimen impact various model sizes differently?
    \item[\textbf{RQ V.}]  Do augmented models transfer better than their non-augmented counterparts?
\end{itemize}

\subsection{In-Domain Experiments}
\label{sec:in-domain-exp}

\begin{figure*}
    \centering
      \includegraphics[width=1\linewidth]{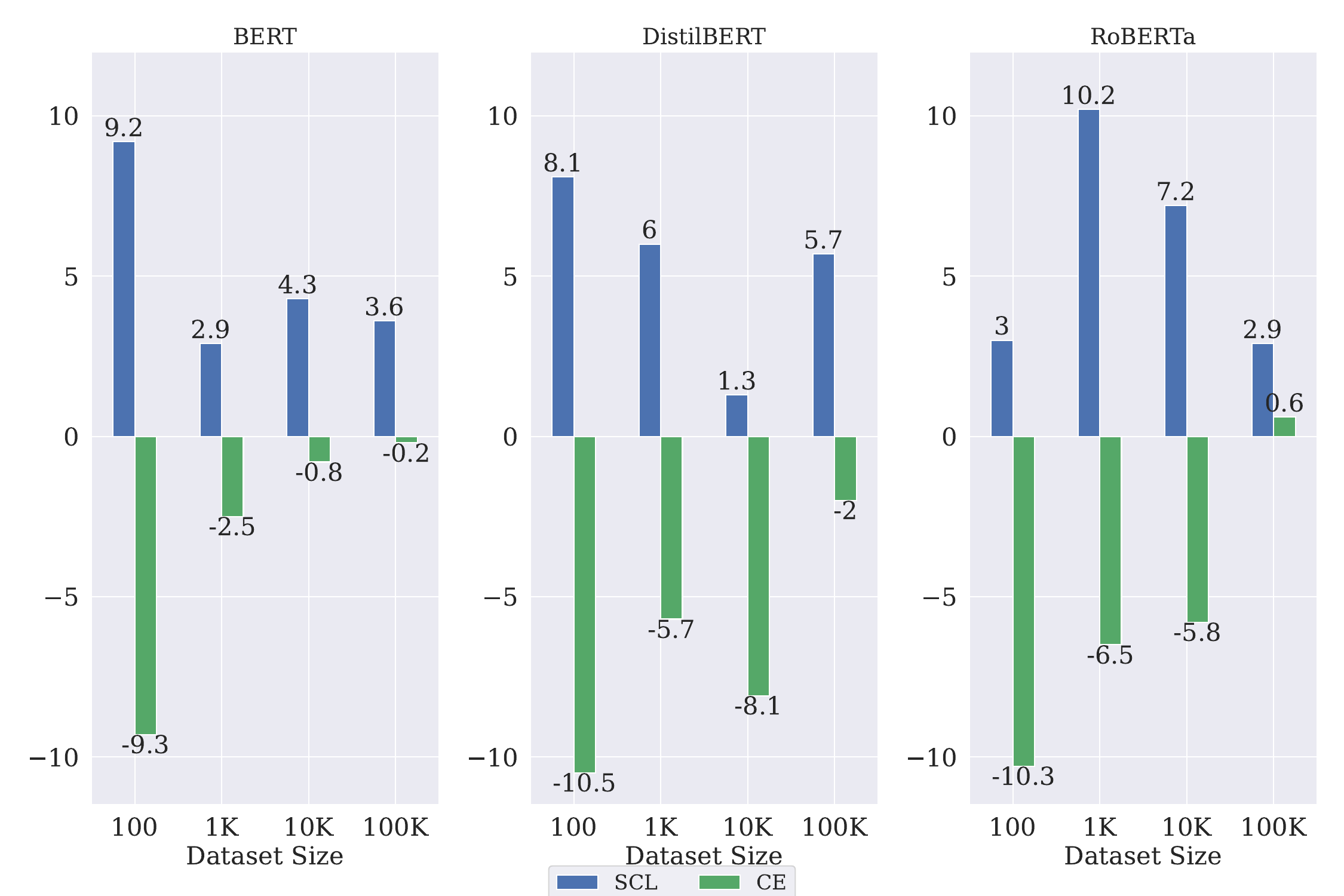}
    \caption{Relative nDCG@10 improvement of \textbf{CE} (cross entropy) model and \textbf{SCL} model over \textbf{Baseline} model. CE model is trained on an augmented dataset using pointwise loss, SCL is trained on an augmented dataset with \ce{}\scl{} loss and the Baseline model is trained on a non-augmented dataset using pointwise loss. The dataset used here is \trecdldf{} and \bm{} augmentation strategy.}
    \label{fig:ce-scl}
\end{figure*}

\begin{table*}
    \centering
    \begin{tabular}{lccccc}
        \toprule
            & \multicolumn{2}{c}{\textbf{Supervised Augmentation}} && \multicolumn{2}{c}{\textbf{Unsupervised Augmentation}} \\
            \cmidrule(lr){2-3}
            \cmidrule(lr){5-6}
            \textit{Ranking Models}
            & $\text{Linear}$ & $\text{Attention}$ && $\text{\bm}$ & $\text{\glove{}}$ \\
        \midrule
        \multicolumn{5}{l}{\bf \bert{}} \\

100 & \textbf{0.599\up{29.3}}$^{*}$ &	0.569\up{22.7}$^{*}$	&& 0.506\up{9.2}$^{*}$	& 0.515\up{11.1}$^{*}$ \\
1k & \underline{0.554}\up{3.7} &   \textbf{0.583}\up{9.1}$^{*}$ &&   0.541\up{2.9}$^{\#}$ &   0.538\up{2.4}$^{\#}$ \\
2k & \underline{0.592}\up{5.1} &   \textbf{0.597}\up{5.9} &&   0.561\up{3.8} &   0.563\up{4}\\
10k & 0.600\up{1.4} & 0.599\up{1.3}$^{\#}$ &&   \underline{0.617}\up{4.3} & \textbf{0.626}\up{5.8}\\
100k & \textbf{0.628}\up{2.2} &   \underline{0.618}\up{0.5} &&  0.602\up{3.6} &   0.611\up{5.1}\\
        \midrule
        \multicolumn{5}{l}{\bf \roberta{}} \\

100 & 0.272\up{15.8}$^{*}$	& \textbf{0.296\up{25.9}}$^{*}$	&&0.242\up{3}	& 0.266\up{13.3} \\
1k & 0.288\down{-3} & 0.296\down{-0.5} &&  \textbf{0.325}\up{10.2} &  \underline{0.308}\up{4.3}\\
2k & \underline{0.516}\up{75.6}$^{*}$ &  \textbf{0.529}\up{80}$^{*}$ && 0.328\up{7.2} &   0.356\up{16.3}\\
10k &  \textbf{0.599}\up{7.6}$^{*}$ &   0.593\up{6.4}$^{*}$ &&   \underline{0.597}\up{7.2}$^{*}$ &   0.596\up{7.0}$^{*}$\\
100k & \textbf{0.637}\up{10.2} &  \underline{0.610}\up{5.6} && 0.598\up{2.9} &   \underline{0.611}\up{5.0}\\
        \midrule
        \multicolumn{5}{l}{\bf \distilbert{}} \\     

100 & 0.247\up{20.7}$^{*}$	& \textbf{0.258\up{25.8}}$^{\#}$	&& 0.222\up{8.1}$^{\#}$	& 0.219\up{7} \\
1k & \underline{0.287}\up{31}$^{*}$ & \textbf{0.372}\up{70}$^{*}$ && 0.254\up{6} & 0.253\up{5.3}\\
2k & \textbf{0.517}\up{85.7}$^{*}$ &  \textbf{0.517}\up{85.7}$^{*}$ &&  0.309\up{6.2}$^{\#}$ &   0.314\up{7.7}\\
10k &  0.545\down{-3.6} &  \underline{0.573}\up{1.4} &&   \underline{0.573\up{1.3}} &   \textbf{0.583}\up{3.1}\\
100k & 0.582\down{-4} &   0.590\down{-2.6} &&   \textbf{0.641}\up{5.7} &   \underline{0.608}\up{0.2} \\
        
        \bottomrule
    \end{tabular}

    \caption{Document re-ranking results on the \trecdldf{} datasets for \ce{} with \scl{} on supervised(Linear, Attention) vs un-supervised (\bm,\glove{}) augmentation. We show the relative improvement of the augmentation approaches against a baseline without augmentation in parentheses. Statistically significant improvements at a level of $95\%$ and $90\%$ are indicated by $*$ and $\#$ respectively~\cite{paired_significance_test}. The best results for each dataset and each model is in \textbf{bold} and second is \underline{underlined} .}
    \label{tab:sup-vs-unsup-augmentation}

\end{table*}

\begin{itemize}
    \item[\textbf{RQ I.}]  Does data augmentation require a different training paradigm?
\end{itemize}
\label{part:rq1}

We first verify the impact of our proposed approach by comparing 3 types of models -- a model with data augmentation and SCL loss (SCL) versus a model trained with only \ce{} loss (baseline) versus a model trained with \ce{} loss and augmented data (CE). We also repeated the experiment with a pairwise objective and found the same trends. The SCL model is trained with augmented data using the \ce{}\scl{} loss objective, i.e. pointwise ranking loss interpolated with the SCL objective. In Figure~\ref{fig:ce-scl} we plot the relative improvement (in terms of nDCG@10) of the CE and \ce{}\scl{} model over the baseline (zero line) (trained without data augmentation) with BM25 augmentation and pointwise objective. We see that only using \ce{} loss with augmented data performs worse than baseline. We find that \ce{}\scl{} more effectively utilizes augmented data to learn better representations which are reflected in consistent improvements over the baseline. As the size of the training set increases, the detrimental effect of data augmentation on the standard CE loss objective diminishes but does not increase sample efficiency. The \ce{}\scl{} model on the other hand sees gains in efficiency for all models. Augmenting a dataset of 100 and 1K examples leads to a 8.1\% and 6\% improvement over the baseline respectively for \distilbert{} only when using our \scl{} loss. This establishes that using data augmentation with traditional ranking loss functions is detrimental to ranking performance. 

\paragraph{\textbf{Insight 1.}} A different training paradigm is required to take advantage of the data augmentation techniques we propose. In particular, adding a contrastive learning objective leads to gains between 1.3\% and 10.2\% in terms of sample efficiency whereas training without an additional objective and the augmented dataset results in lowered sample efficiency.

\begin{table*}
    \centering
    \begin{tabularx}{\textwidth}{rX}
        \toprule
        \textbf{Method} & \textbf{Augmented Document} \\
        \midrule
        \midrule
        \multicolumn{2}{l}{\textbf{Query}: 293327 --- \textit{how many players are in a cricket team} \hfill \textbf{Document}: D863798} \\
        \midrule
        \bm{} & The quota system was introduced as part of South Africa's re-admission into international cricket [\dots] \\
        \cmidrule{2-2}
        \glove{} & South Africa have announced a commitment to a minimum of five black players in their squad for the World Cup [\dots] \\
        \cmidrule{2-2}
        \linear & Salute Michael Jordan at 40Is this the end for Warne? \\
        \cmidrule{2-2}
        \attention{} & A team at any level should be based on the best 11 players available, not on what colour you are. \\
        \midrule
        \midrule
        \multicolumn{2}{l}{\textbf{Query}: 428365 --- \textit{is there a natural muscle relaxer} \hfill \textbf{Document}: D101612} \\
        \midrule
        \bm{} & Which muscle relaxer is best for long term use?. What muscle relaxer [\dots] and muscle spasms chronically. Any non addictive muscle relaxant? [\dots] \\
        \cmidrule{2-2}
        \glove{} & Non habit-forming muscle relaxers Public Forum Discussions Which muscle relaxer is best for long term use? [\dots] \\
        \cmidrule{2-2}
        \linear & is a muscle relaxer an opiate? \\
        \cmidrule{2-2}
        \attention{} & Is Ativan a good muscle relaxant? \\
        \midrule
        \midrule
        \multicolumn{2}{l}{\textbf{Query}: 197720 --- \textit{gussie's house of flowers madison ga} \hfill \textbf{Document}: D3201531} \\
        \midrule
        \bm{} & The idea was simple – he figured he could sell a few hundred floral bouquets to parents for a local high school graduation ceremony [\dots] \\
        \cmidrule{2-2}
        \glove{} & Gussie's House Of Flowers Gussie's House Of Flowers Business name: Gussie's House Of Flowers Address: 136 W Jefferson St City: Madison State: Georgia [\dots] \\
        \cmidrule{2-2}
        \linear & Gussie's House Of Flowers "Gussie's House Of Flowers Business name: Gussie's House Of Flowers Address: 136 W Jefferson St City: Madison State: Georgia \\
        \cmidrule{2-2}
        \attention{} & The block between St. Clair and Madison, which Cason's house is on, is lined with potted fake flowers. \\
        \bottomrule
    \end{tabularx}
    \caption{Anecdotal augmented training instances from \trecdldf{}. \bm{} and \glove{} are unsupervised methods that select paragraphs from the original document denoted here by it's ID. \linear{} and \attention{} are supervised techniques that select relevant sentences based on the query from the same document.}
    \label{tab:aug_examples}
\end{table*}

\begin{itemize}
    \item[\textbf{RQ II.}]  Which data augmentation technique provides the highest sample efficiency gains?
\end{itemize}
 
To answer RQ II, we extend the previous experiment with more data augmentation techniques. In Table~\ref{tab:sup-vs-unsup-augmentation} we compare 2 different classes of augmentation techniques: data augmentation using supervised methods (\linear{}, \attention{}) and unsupervised methods (\bm{}, \glove{}). We use nDCG@10 as the key performance indicator. We trained various models using the \scl{} loss objective on different dataset sizes and different contextual models (\bert{}, \roberta{}, \distilbert{}). 

Our first major observation is that augmentation helps across the board. As dataset size increases, all methods show steady improvement which is not surprising but the magnitude of relative improvement over the baseline is large in nearly all cases. With dataset of 100 instances attention based approach performs 20\% better than baseline on all models. With 1k samples to train on, attention based augmentation with \scl{} leads to a near 70\% relative improvement for \distilbert{}. \bert{} achieves the highest absolute performance on the 1k and 2k samples using attention based augmentation. The exact choice of augmentation technique is not always clear although supervised techniques tend to generally outperform unsupervised techniques. Supervised techniques benefit from being exposed to a ranking oriented sentence selection task beforehand. 

For \roberta{} in particular, we see a different technique winning for each dataset. However, when observing the absolute ndcg@10 measurements, we find that in all cases except for 1k, supervised methods are either comparable or better by a large margin. Unsupervised methods, while simpler, are particularly poor for lean models like \distilbert{}. We believe that supervised methods are usually more capable of constructing meaningful augmentations. Unsupervised methods create noisy samples (see Table~\ref{tab:aug_examples}) by either ignoring semantics (\bm{} -- syntactic matching) or not being aware of the ranking task (\glove{}).

We also observe that the relative improvements over baseline are greater in the case of smaller datasets compared to the larger datasets (which ratifies the findings from \cite{anand2022supervised}) for both classes of approaches. In the case of \roberta{} and \distilbert{} we see an 80\% improvement (statistically significant) over the baseline for the 2K dataset when using supervised augmentation. For the 100k dataset, surprisingly unsupervised methods are key to achieving sample efficiency for \distilbert{} which requires further investigations into the interplay between model size and the ability to learn effectively from augmented data.  

The difference between linear and attention is more evident when observing smaller datasets (100, 1k and 2k). For \bert{} and \distilbert{} we see large sample efficiency gains when using supervised approaches for the 1k dataset but attention results in the largest gains: 3.7\% vs 9.1\% for \bert{} and 31\% vs 70\% for \distilbert{} when compared to linear. The same can be observed for 100 instances where \attention{} gains are 22.7\% vs 29.3\% in case of \bert{} and 25.8\% vs 20.7\% in case of \distilbert{} in comparison with \linear{}

\paragraph{\textbf{Insight 2.}} Supervised augmentation techniques, specifically \attention{} leads to higher sample efficiency rates especially for smaller datasets (100, 1K, 2K) when training with \scl{}.
 
\begin{itemize}
    \item[\textbf{RQ III.}]  Which contrastive loss objective leads to the highest gains when data augmentation type and budget are fixed?
\end{itemize}

\begin{table*}
    \centering
    \begin{tabular}{lllll}
        \toprule
            \textit{\textbf{Losses}}
            & $\textbf{\scl{}}$ & $\textbf{\nce{}}$ & $\textbf{\triplet{}}$ & $\textbf{\nca{}}$ \\
        \midrule
        \multicolumn{5}{l}{\bf \bert{}} \\

100 & 0.569\up{22.7}$^{*}$ &\textbf{0.573\up{23.7}}$^{*}$	& 0.521\up{12.3}$^{*}$	& 0.563\up{21.5}$^{*}$ \\
1k & \textbf{0.583}\up{9.1}$^{*}$ & 0.529\down{-1} & 0.576\up{7.9}$^{*}$ & 0.564\up{5.5} \\
2k & 0.597\up{5.9} & 0.590\up{4.7} & \textbf{0.601}\up{6.6} & 0.580\up{2.9} \\ 
10k & 0.599\up{1.3}$^{\#}$ & 0.603\up{1.9} & \textbf{0.620}\up{4.8} & 0.590\down{-0.4} \\
100k & 0.618\up{0.5} & 0.612\down{-0.5} & \textbf{0.634}\up{3.1} & 0.615\up{0} \\
        \midrule
        \multicolumn{5}{l}{\bf \roberta{}} \\

100 & \textbf{0.296\up{25.9}} & 0.269\up{14.7}	&0.275\up{17.1}$^{*}$	&0.285\up{21.4}$^{*}$ \\
1k & 0.296\down{-0.5} & 0.303\up{1.9} & \textbf{0.317}\up{6.6}$^{*}$ & 0.289\down{-2.9}$^{*}$ \\
2k & 0.529\up{80}$^{*}$ & 0.502\up{70.6}$^{*}$ & \textbf{0.548}\up{86.6}$^{*}$ & 0.534\up{81.7}$^{*}$ \\
10k & 0.593\up{6.4}$^{*}$ & 0.572\up{2.6} & \textbf{0.612}\up{9.9}$^{*}$ & 0.603\up{8.3} \\
100k & 0.610\up{5.6} & 0.602\up{4.1} & \textbf{0.620}\up{7.3} & 0.610\up{5.5} \\
        \midrule
        \multicolumn{5}{l}{\bf \distilbert{}} \\

100 & \textbf{0.258\up{25.8}}$^{\#}$	& 0.235\up{14.5}	& 0.241\up{17.6}	& 0.212\up{3.37}\\
1k & 0.372\up{70}$^{*}$ & 0.368\up{68.1}$^{*}$ & \textbf{0.401}\up{83}$^{*}$ & 0.236\up{7.9} \\
2k & 0.517\up{85.7}$^{*}$ & 0.505\up{81.5}$^{*}$ & \textbf{0.532}\up{91.1}$^{*}$ & 0.485\up{74.3}$^{*}$ \\
10k & 0.573\up{1.4} & 0.563\down{-0.4} & \textbf{0.582}\up{2.9} & 0.551\down{-2.6} \\
100k & 0.590\down{-2.6} & 0.610\up{0.6} & \textbf{0.630}\up{3.3} & 0.561\down{-7.5} \\
        
        \bottomrule
    \end{tabular}
    \caption{nDCG@10 performance of different language models (\bert{}, \roberta{}, and \distilbert{}) on different loss functions (\scl, \nce{}, \triplet{}, and \nca{}) at different training set sizes (100, 1k, 2k, 10k, and 100k) with \textbf{\attention{}} data augmentation. Statistically significant improvements at a level of $95\%$ and $90\%$ are indicated by $*$ and $\#$ respectively~\cite{paired_significance_test}.The best results for each dataset and each model is in bold.}
    \label{tab:point-attention-all-loss-ndcg}
\end{table*}

% Experimental results explained
Different contrastive loss functions have different properties that may make them more or less effective for our particular problem. When the data augmentation type and training budget are fixed, we find that the contrastive loss objective which leads to the highest gains depends on specific characteristics of the dataset and the model being used. To this extent we experiment with 4 different contrastive losses (\triplet{}, \nca{}, \nce{}, \scl{}) with fixed supervised augmentation techniques (\attention{}, \linear{}) for different models (\bert{}, \roberta{}, \distilbert{}). The results are shown in Table~\ref{tab:point-attention-all-loss-ndcg} and Appendix Table~\ref{tab:point-linear-all-loss-ndcg}.

In Table~\ref{tab:point-attention-all-loss-ndcg} we compare different loss objectives while using attention augmentation. It is clear that \triplet{} outperforms all other losses across varying dataset sizes. It shows more than 85\% improvement over baseline (95\% statistically significant) in the case of \roberta{} and \distilbert{} when trained on the 2K dataset. Additionally, the performance of \distilbert{} matches the performance of \bert{} and \roberta{} when training with the 100K dataset. Notice that for \roberta{}, the choice of the loss function is crucial to sample efficiency on the 1k dataset -- with \triplet{} we see a 6.6\% improvement.  

The results also show that increasing the training set size generally leads to better performance across all models and loss functions. However, the rate of improvement varies depending on the model and loss function. For example, for \roberta{}, \triplet{} shows the largest percentage change when increasing the training set size from 1k to 2k, while for \distilbert{}, \nca{} shows the largest percentage change.

\triplet{} loss has 2 unique properties -- the margin hyperparameter and the usage of centroids rather than the actual data points. Max margin based losses are commonly used for ranking tasks~\cite{agarwal2010maximum} which makes this style of contrastive loss more suited to our problem. Additionally, since the augmentation techniques can be noisy, the usage of the centroid dampens it by averaging over all augmentations of the anchor. \triplet{} loss is generally used when there is more intra-class variation which is the case in document ranking where positive documents can vary drastically based on the query at hand. Other loss functions are more suited to classification problems and do not translate as well to our problem setting.

Our empirical findings suggest that pairing data augmentation with the right loss function is imperative to maximizing sample efficiency. Comparing \scl{} loss to \triplet{} we see consistent gains across models and data sizes in terms of nDCG@10 and relative improvement to the baseline. Even for 100k data points, we see gains of 3.1\% to 7.3\% when using \triplet{} which is considerably higher than all other losses.

% insight
\paragraph{\textbf{Insight 3.}} Empirically Centroid Triplet Loss (\triplet) is the superior choice for all models for all dataset sizes when using attention based augmentation.

\begin{itemize}
    \item[\textbf{RQ IV.}]  Does our training regimen impact various model sizes differently?
\end{itemize}

\begin{figure}
    \centering
    \includegraphics[width=0.9\linewidth]{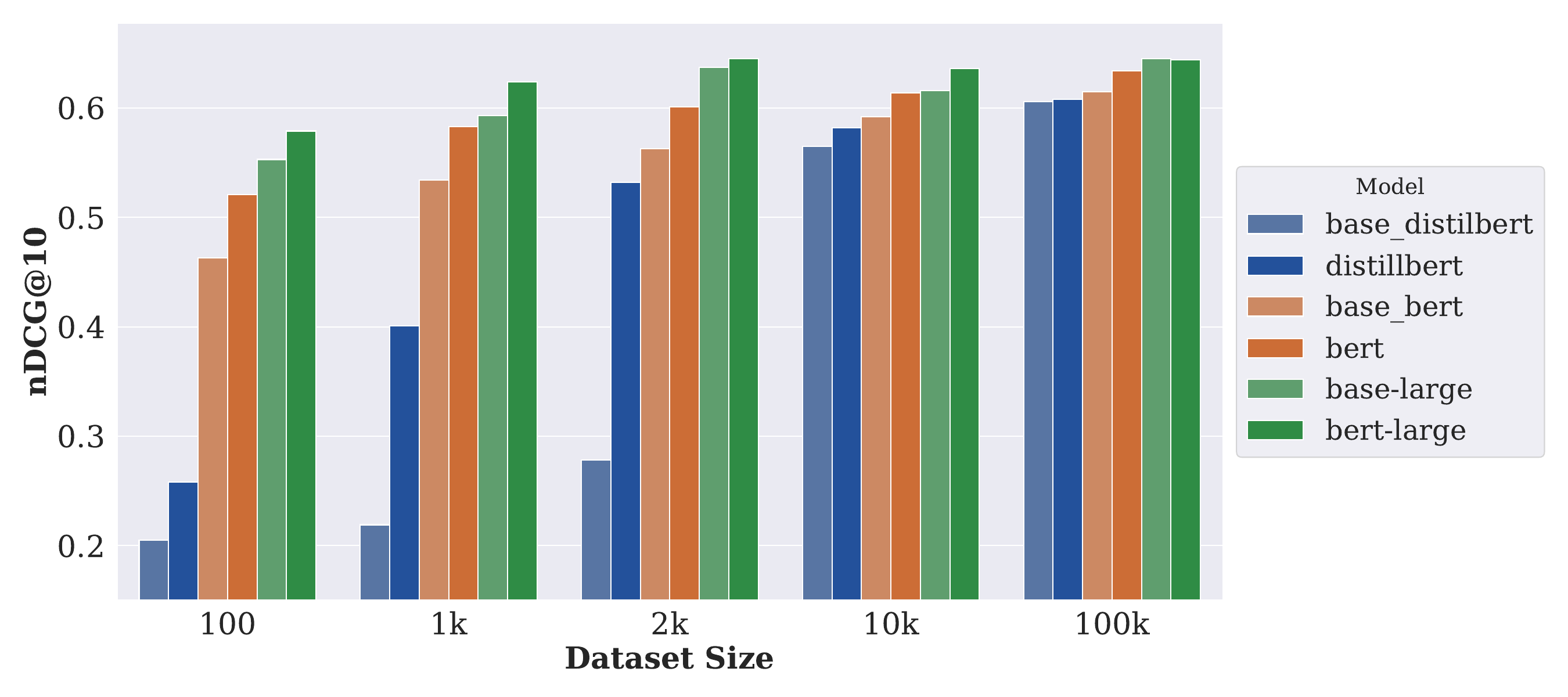}
    \caption{Performance of models of different model sizes compared to their respective baselines. The values with "\textbf{base\_}" represents the base model.}
    \label{fig:diff_model_size}
\end{figure}

% experimental setup
To look at the impact of model sizes on ranking performance we conduct baseline and augmented data experiments on \distilbert{}, \bert{} and \bertl{} by varying the dataset sizes (100, 1K, 2K, 10K, 100K). For the baseline, we use non-augmented data same as all experiments above. For augmented experiments, we use \attention{} augmentation with \triplet{} loss. The performance results (nDCG@10) results are shown in Figure ~\ref{fig:diff_model_size}.

% observations
We observe that larger \bert{}-based language models tend to perform better on smaller datasets, indicating that increasing the model size can compensate for the limited amount of data. Furthermore, we already showed, augmented models outperform their non-augmented counterparts (Insight 1). However, it is worth noting that as the dataset size increases, the performance gap between smaller and larger models narrows, implying that increasing model size may have diminishing returns in the presence of large datasets.

For \distilbert{}, we see that data augmentation leads to large improvements. For practitioners with limited computing budgets, data augmentation provides a significant method to improve the performance of leaner models. Fine tuning lean models on smaller ranking datasets is not always straightforward. We experimented with several hyperparameters to improve the performance of the \distilbert{} baseline for 1k and 2k. However, with augmentation minimal hyperparameter tuning led to gains of over 50\% in sample efficiency. Hard-to-tune models can benefit directly from our approach due to more informative samples and a better loss objective.

%Tables and Figures

% insight
\paragraph{\textbf{Insight 4.}} Larger models (\bertl{}) out perform smaller models (\bert{}, \distilbert{}) in the case of smaller datasets (100, 1k, 2k, 10k) with augmentation primarily benefiting the smallest model (\distilbert{}).

\subsection{Out-of-domain (\beir{}) Experiments }
\label{sec:out-domain-exp}
\begin{itemize}
    \item[\textbf{RQ V.}]  Do augmented models transfer better than their non-augmented counterparts?
\end{itemize}

%Tables and Figures
\begin{table*}
    \centering
    \begin{tabular}{llllll}
        \toprule
            \textit{\textbf{Datasets}}
            & $\textbf{\bert{}}$ & $\textbf{\roberta{}}$ & $\textbf{\distilbert{}}$ & $\textbf{SPLADE}$  \\
        \midrule

SciFact & 0.678\up{3.1} & 0.676\up{134}$^{*}$ & 0.526\up{14.2}$^{*}$ & \textbf{0.699} \\
\quad \hide{Baseline} & \hide{0.658} & \hide{0.288} & \hide{0.461} &    & \\ 
\hline
FiQA & 0.315\up{4.4}$^{*}$ & 0.339\up{50.6}$^{*}$ & 0.211\up{20.5}$^{*}$ & \textbf{0.351} \\
\quad \hide{Baseline} & \hide{0.302} & \hide{0.225} & \hide{0.108} &  \\
\hline

DBPedia & 0.481\down{-0.4}$^{*}$ & \textbf{0.514}\up{36.3}$^{*}$ & 0.435\up{16.2}$^{*}$ & 0.442 \\
\quad \hide{Baseline} & \hide{0.483} & \hide{0.377} & \hide{0.236} &  \\
\hline

TREC-COVID & 0.694\up{2.7}$^{*}$ & \textbf{0.721}\up{21.8}$^{*}$ & 0.603\up{2.3}$^{*}$ & 0.711 \\
\quad \hide{Baseline} & \hide{0.676} & \hide{0.592} & \hide{0.589} &  \\
\hline

NFCorpus & 0.260\up{0.9} & 0.282\up{18.8}$^{*}$ & 0.260\up{2.2}$^{*}$ & \textbf{0.345} \\
\quad \hide{Baseline}  & \hide{0.257} & \hide{0.237} & \hide{0.255} &  \\
\hline

Robust04 & 0.442\up{8.8}$^{\#}$ & 0.429\up{30.3}$^{*}$ & 0.386\up{17}$^{*}$ & \textbf{0.458} \\
\quad \hide{Baseline} & \hide{0.406} & \hide{0.329} & \hide{0.330} &  \\
\hline
\textbf{Average} &  0.478 & 0.494 & 0.404 & \textbf{0.501} \\
        
        \bottomrule
    \end{tabular}
    \caption{nDCG@10 values for different models (\bert{}, \roberta{}, \distilbert{}) on the BEIR dataset for evaluating Out-of-distribution performance. The model used for zero-shot performance are trained on 100K \attention{} dataset with \triplet{} loss. Statistically significant improvements at a level of $95\%$ and $90\%$ are indicated by $*$ and $\#$ respectively~\cite{paired_significance_test}.The best results for each
dataset and each model is in bold.}
    \label{tab:beir-all-sphere-100k}

\end{table*}

% Experimental results explained
Zero-shot out-of-domain transfer for ranking datasets is challenging due to the diversity of topical domains and the information needs of users. Table~\ref{tab:beir-all-sphere-100k} shows the performance of three different \bert{}-based models (\bert{}, \roberta{}, and \distilbert{}) we trained using our proposed paradigm on 6 different datasets against the current SOTA model (SPLADE~\cite{formal2021splade}) on the \beir{} benchmark without any further fine-tuning on each dataset.

We find that our augmented models are significantly better than their corresponding baselines making them not only sample efficient but also more robust. \roberta{} in particular sees large gains across all datasets and at times outperforms SOTA (DBPedia and TREC-COVID). Augmentation helps improve the model's ranking performance more broadly by exposing it to a wider range of examples and scenarios during training. More concretely, we believe that without augmentation, models suffer from the problem of over-fitting which can occur especially when a model is trained on a limited set of examples. Augmentation can help prevent over-fitting by providing the model with a larger and more diverse set of examples to learn from. Furthermore, the added loss objective also helps to regularize the model. 

The \beir{} datasets also varied in the type of queries -- our training dataset \trecdldf{} has simple factual questions whereas DBPedia for instance has entity specific attribute queries and Robust has topical keywords as queries (Table~\ref{tab:query_examples} shows anecdotal evidence).

\begin{table*}
    \centering
    \begin{tabularx}{\linewidth}{ c c X}%{ccccccc}
    \toprule
    \textbf{Dataset} & \textbf{Domain} & \textbf{Queries} \\
    \midrule
    Robust04 & \texttt{News} & price fixing, Russian food crisis, ADD diagnosis treatment, Modern Slavery \\
    FiQA & \texttt{Finance} & How are various types of income taxed differently in the USA?, Looking for good investment vehicle for seasonal work and savings, Understanding the T + 3 settlement days rule, How should I prepare for the next financial crisis?\\
    Dbpedia & \texttt{Entity} & south korean girl groups, electronic music genres, digital music notation formats, FIFA world cup national team winners since 1974 \\
    Trec-Covid & \texttt{Medical} & what is known about those infected with Covid-19 but are asymptomatic?,what evidence is there for the value of hydroxychloroquine in treating Covid-19? \\
    \textsc{NFCorpus} & \texttt{Science} & Is apple cider vinegar good for you?, How can you believe in any scientific study?, organotins, oxen meat\\
    \bottomrule
    \end{tabularx}
    \caption{Example queries from \beir{}. Each dataset varies in topicality and query type. }
    \label{tab:query_examples}
\end{table*}

Our augmentation with contrastive learning exposes the model to more general matching patterns since the model has to not only learn how to estimate relevance between a question and a document but also a sentence and a passage selected by our augmentation. This helps improve the overall semantic understanding of the model which greatly aids transfer. These results hold for all model types which makes a strong case for our augmentation models to be used as universal ranking models when computing and training data is severely limited.

This result shows that we can save costs and increase efficiency by training a single model that performs well across IR datasets with limited amounts of out-of-domain training data. Practitioners need not invest in further fine tuning or deploying individual models for each dataset. 

% insight
\paragraph{\textbf{Insight 5.}} Augmented models are better suited for zero-shot transfer because our approach improves the model's ability to estimate relevance between two pieces of text by training on more diverse examples.

\subsection{Limitations}
\label{sec:fail}

In our experiments, we considered 3 contextual models, 4 losses, 4 augmentation techniques, and 4 dataset sizes. 
We ran experiments with all combinations of dataset size, loss and augmentation. We had 2 key questions to verify our insights from the previous sections:

\begin{itemize}
    \item Is attention based augmentation always the right choice irrespective of the loss function?
    \item Is \triplet{} loss the best choice irrespective of augmentation technique?
\end{itemize}

\begin{figure}
    \centering
    \includegraphics[width=1\linewidth]{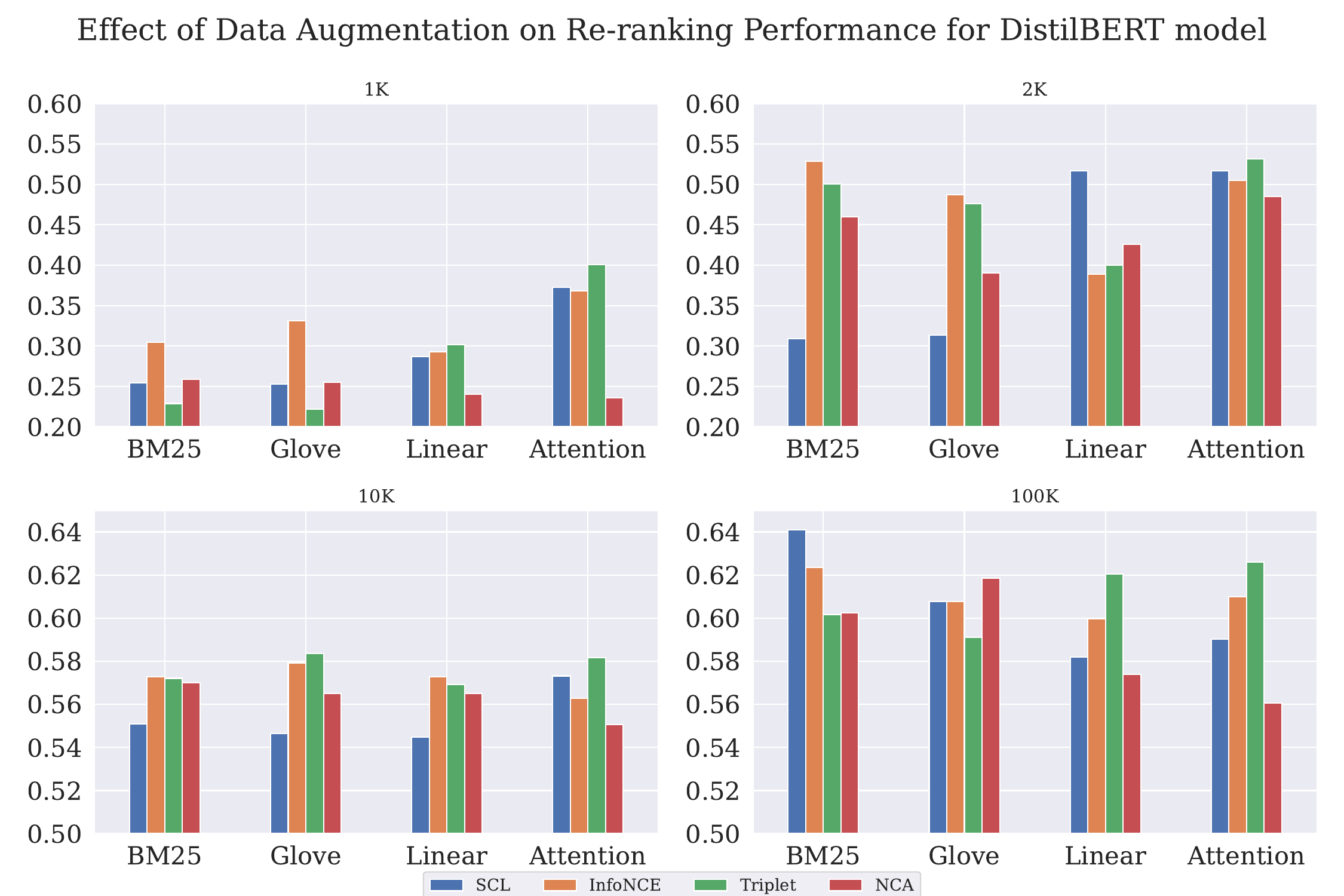}
    \caption{\distilbert{} nDCG@10 performance on difference size datasets with different augmentation and losses.}
    \label{fig:distilbert_all_results_ms}
\end{figure}

From Figure~\ref{fig:distilbert_all_results_ms} it is apparent that attention augmentation and \triplet{} are not always the best choice. Even though the best performance at 1k, 2k and 10k is from the proposed combination, we see certain combinations being close to or surpassing it in the case of 100k (\bm{} + \scl{}). 

In the case of 1k and 2k, \triplet{} performs considerably better than all other losses for unsupervised augmentation. When the dataset size increases however, the differences are much smaller. When using a 100K dataset, the choice of the loss function and augmentation is not clear. \scl{} performs relatively poorly ,especially for unsupervised augmentation in 1k and 2k but slightly outperforms our proposed combination for 100k. \triplet{} does not result in the best performance in low data regimes when using unsupervised augmentation.

In the previous section, we studied the impact of various losses on attention augmentation. Attention and linear augmentation techniques are both supervised selectors from~\cite{leonhardt2021learnt}. In both cases, the objective of the selector is to pick the most relevant sentence. The linear selector pools word embeddings using a max operation so positional context is lost. This design choice leads to attention
outperforming linear irrespective of the loss function and dataset size. For 1k and 2k, linear is outperformed by \glove{} and \bm{}. This could mean that selecting noisy sentences for contrastive learning is worse than simple term matching when paired with the correct loss. 
Supervised augmentation alone is not sufficient to gain the best performance for 1k and 2k. It must be paired with the right loss. \triplet{} in general is seemingly a good choice for a loss function

If we consider the simplicity of implementation, the combination of \bm{} and \nce{} is a good choice. It rivals \attention{} and \triplet{} in all datasets above 1k. For 1k however, attention based augmentation makes the largest difference since all losses except \nca{} exhibit large improvements. The value of attention based augmentation compared to other augmentations diminishes as the data size grows. While augmentation still leads to overall sample efficiency gains, the choice of augmentation is not as crucial only if paired with the right loss function. Empirically trying to detect which combination works the best for a practitioner's dataset however goes against the spirit of efficiency. Our proposed combination of \triplet{} and \attention{} displays the clearest trend even if there are a few cases where it is not the best. 
\section{Discussion and Conclusion}
\label{sec:conclu}

In this paper, we empirically explored the impact of data augmentation on sample efficiency for ranking problems. Our central premise is that data augmentation when paired with a contrastive learning objective leads to significant improvements in performance with the same number of training instances. Our experimental space includes both supervised and unsupervised augmentation techniques and 4 different contrastive learning objectives. We found that a combination of attention-based supervised augmentation and \triplet{} provides the highest sample efficiency gains for a range of models and dataset sizes. We see large benefits for smaller models on smaller dataset sizes which is an important step toward their wider adoption. \distilbert{} sees gains of up to 85\% when fine-tuned using our proposed setup. We also observe that not all models exhibit the same level of gains with \bert{} gaining between 1\% and 10\% depending on the size of the dataset.
Another benefit of augmented training is the drastic improvement in zero-shot transfer. We showed that our best-augmented models improve performance by large margins compared to their non-augmented counterparts. \roberta{} on average sees a near 60\% improvement when augmented and can rival fine-tuned SOTA models. 

In conclusion, we believe the approach we propose is only the first step towards sample efficient training of ranking models with contrastive losses. Augmentation and adjusting loss objectives are cheaper alternatives for most practitioners instead of gathering expensive training data. There remain several areas of future work -- for instance, augmentation techniques that can operate on queries is still under-explored. We still lack a clear understanding of the impact of various contrastive losses on the type of augmentation. Further research is needed to identify why a specific loss function benefits from a particular type of augmentation. We are also yet to explore the usage of synthetic training data in this context. Generative models have been used in the past to augment datasets with new queries, which we can also leverage in future work.

\begin{acks}
This work is supported by the European Union – Horizon 2020 Program under the scheme “INFRAIA-01-2018-2019 – Integrating Activities for Advanced Communities”, Grant Agreement n.871042, “SoBigData++: European Integrated Infrastructure for Social Mining and Big Data Analytics” (http://www.sobigdata.eu).

This work is supported in part by the Science and Engineering Research Board, Department of Science and Technology, Government of India, under Project SRG/2022/001548. Koustav Rudra is a recipient of the DST-INSPIRE Faculty Fellowship [DST/INSPIRE/04/2021/003055] in the year 2021 under Engineering Sciences.
\end{acks}

\bibliographystyle{ACM-Reference-Format}
\balance
\bibliography{reference}
\clearpage
\appendix
\section{Appendix}
Here we add additional details relating to experimental setup and also show additional results.
\label{appendix:appendix}
% \section{Experimental Setup}
% \subsection{Experimental Space}
We have a large experimental space. Here are the number models best trained : 5 Datasets * (3 Models * 4 Loss Function * 4 Augmentation Techniques) + (5 Datasets * 3 Models) Baseline = 255 models
\begin{itemize}
    \item Datasets: 100, 1K, 2k, 10K, 100K
    \item Models: \bm{}, \roberta{}, \distilbert{}
    \item Losses: \scl{}, \triplet{}, \nca{}, \nce{}
    \item Augmentation techniques: \bm{}, \glove{}, \attention{}, \linear{}
\end{itemize}

This does not include intermediate models, models trained with different hyperparameters. That would increase the number of models trained by a factor of 5 on average.
\subsection{Hyperparameters}
\label{appendix:hyperparameters}
    \begin{table}[ht]
\centering
\label{tab:learning_rates}
\begin{tabular}{ccl}
\toprule
Learning Rate & Model & Dataset size (\# instances) \\
\midrule
3e-3 & \bert{}, \roberta{}, \distilbert{} & $100$\\
1e-4 & \bert{}, \roberta{} & $1000$; $2000$; $10,000$ \\
1e-5 & \bert{}, \roberta{} & $100,000$ \\
3e-4 & \distilbert{} & $1000$; $2000$; $10,000$ \\
3e-5 & \distilbert{} & $100,000$ \\
\bottomrule
\end{tabular}
\caption{Learning rates used for training models for different datasets}
\vspace{-1cm}
\label{tab:hyperparameter}
\end{table}

%\subsection{Other Experimental details}
%Hardware: Used 1 Nvidia A100 40GB for all models. Only used batch sizes which could only fit a single GPU.

\subsection{Results}
\label{appendix:results}

Here in Table \ref{tab:bert_bm25_full} we compare the performance of full \ms{} document dataset with 100K dataset. Both the datasets are augmented using \bm{} augmentation and the loss used here is \scl{}. We can see that the performance difference between the two datasets is not large. The absolute values start converging as the dataset grows.
\begin{table}
    %\centering
    \begin{tabular}{lll}
        \toprule
            \textit{\textbf{Model}} & $\textbf{full}$ & $\textbf{100K}$\\
        \midrule
\bert{} & 0.648\up{2.4} & 0.602\up{3.6}\\
\roberta{} & 0.652\up{11.2} & 0.598\up{2.9}\\
\distilbert{} & 0.653\up{6.6} & 0.641\up{5.7}\\
        
        \bottomrule
    \end{tabular}
    % }
    \caption{Comparing nDCG@10 values for different models (\bert{}, \roberta{}, \distilbert{}) on the full \ms{} and 100K dataset with \bm{} and \scl{}. The percentage improvements in brackets are improvements from their respective baselines.}
    \label{tab:bert_bm25_full}
\end{table}
\begin{table*}
    \centering
    \begin{tabular}{lllll}
        \toprule
            \textit{\textbf{Losses}}
            & $\textbf{\scl{}}$ & $\textbf{\nce{}}$ & $\textbf{\triplet{}}$ & $\textbf{\nca{}}$  \\
        \midrule
        \multicolumn{5}{l}{\bf \bert{}} \\
1k & 0.554\up{3.8} & \textbf{0.572}\up{7.0} & 0.541\up{1.2} & 0.558\up{4.4} \\
2k & 0.592\up{5} & 0.584\up{3.8} & 0.590\up{4.8} & \textbf{0.603}\up{7.1} \\
10k & 0.600\up{1.4} & 0.607\up{2.7} & \textbf{0.620}\up{4.7} & 0.605\up{2.2} \\
100k & 0.628\up{2.2} & 0.610\down{-0.9} & \textbf{0.637}\up{3.6} & 0.626\up{1.8} \\
        \midrule
        \multicolumn{5}{l}{\bf \roberta{}} \\
1k & \textbf{0.288}\down{-3} & 0.287\down{-3.5}$^{*}$ & 0.271\down{-9}$^{\#}$ & 0.273\down{-8} \\
2k & \textbf{0.516}\up{75.6}$^{*}$ & 0.427\up{45.4}$^{*}$ & 0.421\up{43.3}$^{*}$ & 0.470\up{59.9}$^{*}$ \\
10k & 0.594\up{6.6}$^{*}$ & 0.570\up{2.3} & \textbf{0.609}\up{9.3}$^{*}$ & 0.588\up{5.6}$^{*}$ \\
100k & \textbf{0.637}\up{10.2} & 0.619\up{7.1} & 0.615\up{6.5} & 0.632\up{9.4} \\
        \midrule
        \multicolumn{5}{l}{\bf \distilbert{}} \\
1k & 0.287\up{31.1}$^{*}$ & 0.293\up{33.8}$^{*}$ & \textbf{0.302}\up{37.9}$^{*}$ & 0.240\up{9.7} \\
2k & \textbf{0.517}\up{85.7}$^{*}$ & 0.389\up{39.9}$^{*}$ & 0.400\up{43.8}$^{*}$ & 0.426\up{53.1}$^{*}$ \\
10k & 0.545\down{-3.6} & \textbf{0.573}\up{1.4} & 0.569\up{0.7} & 0.565\up{0.0} \\
100k & 0.582\down{-4} & 0.600\down{-1.1} & \textbf{0.620}\up{2.3} & 0.574\up{5.3} \\
        
        \bottomrule
    \end{tabular}
    \caption{nDCG@10 performance of different language models (\bert{}, \roberta{}, and \distilbert{}) on different loss functions (\scl, \nce{}, \triplet{}, and \nca{}) at different training set sizes (1k, 2k, 10k, and 100k) with \textbf{\linear{}} data augmentation. Statistically significant improvements at a level of $95\%$ and $90\%$ are indicated by $*$ and $\#$ respectively~\cite{paired_significance_test}.The best results for each dataset and each model is in bold}
    \label{tab:point-linear-all-loss-ndcg}
\end{table*}
\end{document}